\documentclass{myaa}

\usepackage[varg]{txfonts}

\usepackage{natbib}
\usepackage{graphicx}
\usepackage{epstopdf}
\usepackage{amssymb}
\usepackage{float}

\begin{document}

\title{Model of Saturn's protoplanetary disk forming in-situ its regular satellites and innermost rings before the planet is formed}

\author{Dimitris M. Christodoulou\inst{1,2}  
\and 
Demosthenes Kazanas\inst{3}
}

\institute{
Lowell Center for Space Science and Technology, University of Massachusetts Lowell, Lowell, MA, 01854, USA.\\
\and
Dept. of Mathematical Sciences, Univ. of Massachusetts Lowell, 
Lowell, MA, 01854, USA. \\ E-mail: dimitris\_christodoulou@uml.edu\\
\and
NASA/GSFC, Laboratory for High-Energy Astrophysics, Code 663, Greenbelt, MD 20771, USA. \\ E-mail: demos.kazanas@nasa.gov \\
}


\def\gsim{\mathrel{\raise.5ex\hbox{$>$}\mkern-14mu
                \lower0.6ex\hbox{$\sim$}}}

\def\lsim{\mathrel{\raise.3ex\hbox{$<$}\mkern-14mu
               \lower0.6ex\hbox{$\sim$}}}

\abstract{
We fit an isothermal oscillatory density model of Saturn's protoplanetary disk to the present-day major satellites and innermost rings D/C and we determine the radial scale length of the disk, the equation of state and the central density of the primordial gas, and the rotational state of the Saturnian nebula. This disk does not look like the Jovian and Uranian disks that we modeled previously. Its power-law index is extremely steep ($k=-4.5$) and its radial extent is very narrow ($\Delta R\lesssim 0.9$ Gm), its rotation parameter that measures centrifugal support against self-gravity is somewhat larger ($\beta_0=0.0431$), as is its radial scale length (395 km); but, as was expected, the size of the Saturnian disk, $R_{\rm max}=3.6$ Gm, takes just an intermediate value. On the other hand, the central density of the compact Saturnian core and its angular velocity are both comparable to that of Jupiter's core (density of $\approx 0.3$~g~cm$^{-3}$ in both cases, and rotation period of 5.0 d versus 6.8 d); and significantly less than the corresponding parameters of Uranus' core. As with the other primordial nebulae, this rotation is sufficiently slow to guarantee the disk's long-term stability against self-gravity induced instabilities for millions of years of evolution.}

\authorrunning{ }
\titlerunning{Formation of the major Saturnian satellites and rings}

\maketitle


\section{Introduction}\label{intro}

In previous work \citep{chr19a,chr19b,chr19c}, we presented isothermal models of the solar, Jovian, and Uranian primordial nebulae capable of forming protoplanets and, respectively, protosatellites long before the central object is actually formed by accretion processes. This entirely new ``bottom-up'' formation scenario is currently observed in real time by the latest high-resolution ($\sim$1-5~AU) observations of many protostellar disks by the ALMA telescope \citep{alm15,and16,rua17,lee17,lee18,mac18,ave18,cla18,kep18,guz18,ise18,zha18,dul18,fav18,har18,hua18,per18,kud18,lon18,pin18,vdm19}.   In this work, we apply the same model to Saturn's primordial disk that formed its eight major satellites and the inner rings D/C. Our goal is to compare our best-fit model of Saturn's primordial nebula to Jupiter and Uranus' nebulae and to find similarities and differences between the three disks that hosted gravitational potential minima in which the orbiting moons could form in relative safety over millions of years of evolution.

The model nebula of Saturn is somewhat similar to that of Jupiter and dissimilar to that of Uranus. Compared to Jupiter's physical parameters, the Saturnian radial scale length is 395 km versus 368 km and the core density is 0.27 g~cm$^{-3}$ versus 0.31 g~cm$^{-3}$. The maximum size of the Saturnian disk, $R_{\rm max}=3.6$ Gm is intermediate to those of the other two protoplanets (12 Gm and 0.60 Gm for Jupiter and Uranus, respectively). In addition, Saturn's uniform core size is slightly larger than Jupiter's ($R_1=0.321$ Gm versus 0.220 Gm). This is necessary because Saturn's core hosts five closely packed density maxima as opposed to Jupiter's core that hosts only two widespread density maxima. On the other hand, the outer flat-density region of Saturn's disk  $R_2=1.21$ Gm is a lot smaller than $R_{2} = 5.37$ Gm of Jupiter's disk. Finally, the Saturnian disk exhibits significantly higher rotational support against self-gravity (Saturn's $\beta_0$ is about 50\% larger than Jupiter's), but still this value is sufficiently low to guarantee long-term dynamical stability. Just as in the case of Jupiter's primordial disk, the high central densities and the mild differential rotation speeds of Saturn's nebula signify that its major equatorial moons and its rings were formed in-situ long before Saturn was actually fully formed.

The analytic (intrinsic) and numerical (oscillatory) solutions of the isothermal Lane-Emden equation and the resulting model of the gaseous nebula have been described in detail in \cite{chr19b,chr19c} for the primordial disks of Jupiter and Uranus and we will not repeat these descriptions here. 
In what follows, we apply in \S~\ref{models2} our model nebula to the major moons and the inner rings D/C of Saturn and we compare the best-fit results to Jupiter's extended {\it Model 2} and Uranus' best-fit model. In \S~\ref{disc}, we summarize and discuss our results.

\begin{figure}
\begin{center}
    \leavevmode
      \includegraphics[trim=0.2 0.2cm 0.2 0.2cm, clip, angle=0, width=10 cm]{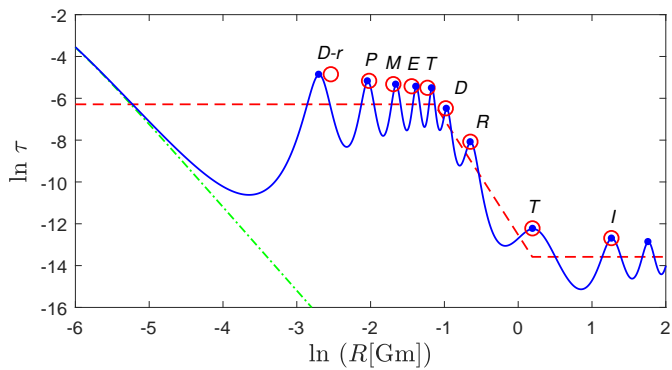}
      \caption{Equilibrium density profile for the midplane of Saturn's primordial protoplanetary disk that formed its rings and largest moons. The center of ring D and the minor moon Pan were also included because they improve the fit considerably. (Key: D-r:D-ring, P:Pan, M:Mimas, E:Enceladus, T:Tethys, D:Dione, R:Rhea, T:Titan, I:Iapetus.) The best-fit parameters are $k=-4.5$, $\beta_0=0.0431$, $R_1=0.321$~Gm, and $R_2=1.21$~Gm. The radial scale length of the disk is $R_0=395$~km. The Cauchy solution (solid line) has been fitted to the present-day moons of Saturn (and the center of the D-ring) so that its density maxima (dots) correspond to the observed semimajor axes of the orbits of the moons (open circles). The density maximum corresponding to the location of Titan was scaled to a distance of $R_T=1.222$~Gm. The mean relative error of the fit is 6.7\%, affirming that this simple equilibrium model produces a good match to the observed data points. The intrinsic solution (dashed line) and the nonrotating analytical solution (dash-dotted line) are also shown for reference. 
\label{fig1}}
  \end{center}
\end{figure}

\section{Physical Model of Saturn's Protoplanetary Disk}\label{models2}

\subsection{Best-Fit Saturnian disk model}\label{model1}

The numerical integrations that produce oscillatory density profiles were performed with the \textsc{Matlab} {\tt ode15s} integrator \citep{sha97,sha99} and the optimization used the Nelder-Mead simplex algorithm as implemented by \cite{lag98}. This method (\textsc{Matlab} routine {\tt fminsearch}) does not use any numerical or analytical gradients in its search procedure which makes it extremely stable numerically, although it is somewhat slow because it proceeds with direct function evaluations only in the multidimensional space defined by the free parameters.

In Fig.~\ref{fig1}, we show the best optimized fit to the semimajor axes of the moons of Saturn, including the innermost D-ring and the small moon Pan. Inclusion of the D-ring is absolutely necessary in order to reduce the errors of the fit to below 20\%. Without the D-ring, the first density maximum of the oscillatory solution cannot get out to the orbit of Pan or to the orbit of any other minor moonlet interior to the orbit of Pan. Ring C lies very closely to ring D and ring D is not assigned to a separate density maximum. Both rings appear to have formed inside the same innermost potential trough that is very wide. Ring B also appears to exist within the same wide trough. Ring A is not used because it is almost coincident with the orbit of Pan, so both of them appear to have formed inside the same potential trough. 

In this modeling, we have used all four available free parameters ($k$, $\beta_0$, $R_1$, and $R_2$) to fit the current orbits of the eight satellites and the D-ring, and the best-fit model turns out to be of good quality (mean relative error of 6.7\%, all of which is coming from the inaccuracy in the position of the D-ring). The inner core parameter $R_1$ is no longer correlated to $\beta_0$ \citep{chr19b,chr19c} and we think this is because the model needs an excessively large core to host 4 moons and the D-ring as well; none of the previous models of disks around solar-system protoplanets required so many moons inside the uniform core. The parameter $R_2$ is also necessary in order to fit simultaneously the orbits of the outer moons Titan and Iapetus.

We find the following physical parameters from the best-fit model: $k=-4.5$, $\beta_0=0.0431$, $R_1=0.321$ Gm (close to the orbit of the inner moon Tethys), and $R_2 = 1.21$ Gm (nearly coincident to the orbit of Saturn's largest moon Titan). The radial scale of the model was determined by fitting the density peak that corresponds to the orbit of Titan to its distance of 1.222 Gm, and the scale length of the disk then turns out to be $R_0=395$ km. The best-fit model is certainly stable to nonaxisymmetric self-gravitating instabilities because of the low value of $\beta_0$ \citep[the critical value for the onset of dynamical instabilities is $\beta_*\simeq 0.50$;][]{chr95}.

The model disk extends out to 7.4 Gm ($\ln R=2$ in Fig.~\ref{fig1}), but its validity ends around the distance of the outermost major moon Iapetus ($R_{\rm max}\approx 3.56$ Gm). The next outer density peak lies at a distance of 5.84 Gm around which no moon is known. The disk of Saturn must have been small ($< 4$-5  Gm in radial extent) because the next outer irregular moon, Kiviuq, has a semimajor axis of 11.3 Gm and Phoebe, the large and most important irregular moon, lies even farther out at 12.9 Gm. The gap between Iapetus and Kiviuq (or Phoebe) is enormous, and no moons or moonlets are found in this region, so it must have been empty from the very beginning of the formation of the system.

\subsection{Physical parameters from the best-fit Saturnian model}\label{rhomax1}

Using the scale length of the disk $R_0$ and the definition $R_0^2 = c_0^2/(4\pi G\rho_0)$, we write the equation of state for the Saturnian circumplanetary gas as
\begin{equation}
\frac{c_0^2}{\rho_0} \ = \ 4\pi G R_0^2 \ = \ 1.31\times 10^{9} 
{\rm ~cm}^5 {\rm ~g}^{-1} {\rm ~s}^{-2}\, ,
\label{crho1}
\end{equation}
where $c_0$ and $\rho_0$ are the local sound speed and the local density in the inner disk, respectively, and $G$ is the gravitational constant.
For an isothermal gas at temperature $T$, ~$c_0^2 = {\cal R} T/\overline{\mu}$, where $\overline{\mu}$ is the mean molecular weight and ${\cal R}$ is the
universal gas constant. Hence, eq.~(\ref{crho1}) can be rewritten as
\begin{equation}
\rho_0 \ = \ 0.0637 \,\left(\frac{T}{\overline{\mu}}\right) \
{\rm ~g} {\rm ~cm}^{-3}\, ,
\label{trho1}
\end{equation}
where $T$ and $\overline{\mu}$ are measured in degrees Kelvin and 
${\rm ~g} {\rm ~mol}^{-1}$, respectively. 

For the coldest gas with $T \geq 10$~K 
and $\overline{\mu} = 2.34 {\rm ~g} {\rm ~mol}^{-1}$ (molecular hydrogen and
neutral helium with fractional abundances $X=0.70$ and $Y=0.28$ by
mass, respectively), we find that
\begin{equation}
\rho_0 \ \geq \ 0.27 \ {\rm ~g} {\rm ~cm}^{-3}\, .
\label{therho1}
\end{equation}
This high value implies that the conditions for protosatellite formation were already in place during the early isothermal phase \citep{toh02} of the Saturnian nebula.

Using the above characteristic density $\rho_0$ of the inner disk
in the definition of ~$\Omega_J\equiv\sqrt{2\pi G\rho_0}$, we determine the Jeans frequency of the disk:
\begin{equation}
\Omega_J \ = \ 3.4\times 10^{-4} {\rm ~rad} {\rm ~s}^{-1}\, .
\label{thej1}
\end{equation}
Then, using the model's value $\beta_0 = 0.0431$ in the definition 
of ~$\beta_0\equiv \Omega_0 /\Omega_J$, we determine the angular velocity of the uniformly-rotating core ($R_1\leq 0.321$~Gm), viz.
\begin{equation}
\Omega_0 \ = \ 1.5\times 10^{-5} {\rm ~rad} {\rm ~s}^{-1}\, .
\label{theom1}
\end{equation}
For reference, this value of $\Omega_0$ for the core of the Uranian nebula corresponds to an orbital period of $P_0=5.0$~d. This value is close to the present-day orbital period of Rhea (4.5 d), but it is not near the orbital period of the largest moon Titan (16 d). This is a deviation from what we found for the solar system and for Jupiter and more consistent with the disk model of Uranus: the angular velocity of the core of the primordial Saturnian nebula is comparable to the present-day angular velocity of the second largest regular satellite Rhea, but it still lands in a region where large moons were formed in the Saturnian nebula. All our protoplanetary models of gaseous-giant disks so far support this property. It remains to be examined in the case of Neptune whose regular satellites are orbiting in a closely packed configuration, challenging all the nebular models constructed so far.

\begin{table*}
\caption{Comparison of the protoplanetary disks of Jupiter, Uranus, and Saturn}
\label{table1}
\begin{tabular}{lllll}
\hline
Property & Property & Jupiter's & Uranus' & Saturn's\\
Name     & Symbol (Unit) & Model 2 & Model& Model \\
\hline
Density power-law index & $k$                                          &   $-1.4$  	     & $-0.96$ &  $-4.5$ \\

Rotational parameter & $\beta_0$                                &    0.0295 	       &  0.00507 &  0.0431  \\

Inner core radius & $R_1$ (Gm)                              &   0.220  	       &  0.0967 &  0.321 \\

Outer flat-density radius & $R_2$ (Gm)                              &   5.37        	   &  $\cdots$  & 1.21  \\

Radial extent of the density power law & $\Delta R$ (Gm) & 5 & $\cdots$ & 0.9  \\

Scale length & $R_0$ (km)                               &   368        	   &  27.6 & 395 \\

Equation of state & $c_0^2/\rho_0$ (${\rm cm}^5 {\rm ~g}^{-1} {\rm ~s}^{-2}$) & $1.14\times 10^9$ & $6.39\times 10^6$ &  $1.31\times 10^9$ \\

Minimum core density for $T=10$~K, $\overline{\mu} = 2.34$ & $\rho_0$ (g~cm$^{-3}$)         &    0.31   			&  55.6  & 0.27 \\

Isothermal sound speed for $T=10$~K, $\overline{\mu} = 2.34$ & $c_0$ (m~s$^{-1}$) & 188 & 188  & 188  \\

Jeans gravitational frequency & $\Omega_J$ (rad~s$^{-1}$)    &    $3.6\times 10^{-4}$ & $4.8\times 10^{-3}$  & $3.4\times 10^{-4}$  \\

Core angular velocity & $\Omega_0$ (rad~s$^{-1}$)    &    $1.1\times 10^{-5}$ 	& $2.5\times 10^{-5}$  &   $1.5\times 10^{-5}$   \\

Core rotation period & $P_0$ (d)                                 &    6.8 	   			&  3.0 &  5.0 \\

Maximum disk size & $R_{\rm max}$ (Gm)                &    12 	   			&   0.60  & 3.6 \\
\hline
\end{tabular}
\end{table*}

\subsection{Comparison between all best-fit models}\label{comp}

We show a comparison between the physical parameters of the best-fit models of Saturn, Uranus, and Jupiter in Table~\ref{table1}. It is obvious that the Saturnian nebula shares more common characteristics with the Jovian nebula than with the Uranian nebula. This is not surprising, given the present-day size and orbit of Saturn. Despite being 3.3 times smaller ($R_{\rm max}$), the primordial disk of Saturn appears to be about as heavy as the Jovian disk ($\Omega_J$ and $\rho_0$) and just about as compact ($R_0$).
Furthermore, the inner uniform core ($R_1$) appears to rotate slower than the extremely compact core of Uranus at an angular velocity ($\Omega_0$) very much comparable to that of the Jovian core. 

The power-law index of the Saturnian nebular model is $k\approx -4.5$ (surface density $\Sigma\propto R^{-4.5}$), much steeper than the other two nebular models. Such an extreme value of $k$ has never been observed in studies of young circumstellar disks in the pre-ALMA era \citep[][and references within]{and07,hun10,lee18}. We believe that this could be a characteristic of some protoplanetary disks only, whereas large-scale protostellar disks should not exhibit such extremely steep density profiles.
Furthermore, the Saturnian disk hosts high enough densities to ensure that a ``bottom-up'' hierachical formation occurred around this protoplanet as well. As we have found for the other gaseous giants in the solar system, protosatellites are seeded early inside their nebular disks and long before the protoplanets are fully formed; these compact moon/ring systems come to be in $< 0.1$ Myr \citep{har18} and long before the central star becomes fully formed \citep[see also][]{gre10}.

\section{Summary and Discussion}\label{disc}

We have constructed isothermal differentially-rotating protoplanetary models of the Saturnian nebula, the primordial disk in which the regular moons and inner rings were formed (\S~\ref{models2}). The best-fit model is shown in Fig.~\ref{fig1} and its physical parameters are listed in Table~\ref{table1}. In the optimization, we retained also the innermost ring D and the minor moon Pan (nearly coincident with ring A) in order to fit the two density maxima near the center that form deep inside the large uniform core of the model. These additions allowed us to find a much better model for the seven major moons of Saturn. The mean relative error of 6.7\% in the best-fit model stems entirely from the inaccuracy in the position of the D-ring.

We have compared this model to the best-fit models of Jupiter and Uranus \citep{chr19b,chr19c} (\S~\ref{comp}). Saturn's disk appears to be closer to the larger disk of Jupiter and very different than the smaller disk of Uranus; although it also exhibits a unique feature, an extremely steep power-law index $k=-4.5$ extending over a very narrow region of size $\Delta R\approx 0.9$ Gm. All of these models appear to be stable and long-lived, so it seems that their regular moons and ring structures could form early in the evolution of each nebula and long before the protoplanets managed to pull their gaseous envelopes on to their solid cores. Once again, the results support strongly a ``bottom-up'' scenario in which regular satellites form first, followed by their planets, and then by the central star.

Neptune's regular moons have an arrangement that is very different than the moons of the gaseous giants that we modeled so far \citep{jac04}. The six regular moons cover an annular region of only 0.07 Gm, and this extremely compact configuration represents a challenge in modeling its primordial disk. This challenge will be addressed in a forthcoming modeling effort.

\end{document}